\documentclass[twocolumn,prc,aps,showpacs,superscriptaddress,amsmath,amssymb,floatfix,nofootinbib,superscriptaddress]{revtex4}
\usepackage{graphicx}
\usepackage{epsfig}

\newcommand{\fpi}{\mbox{$F_\pi$}}
\newcommand{\qsq}{\mbox{$Q^2$}}
\newcommand{\sigl}{\mbox{$\sigma_{\mathrm{L}}$}}
\newcommand{\sigt}{\mbox{$\sigma_{\mathrm{T}}$}}
\newcommand{\siglt}{\mbox{$\sigma_{\mathrm{LT}}$}}
\newcommand{\sigtt}{\mbox{$\sigma_{\mathrm{TT}}$}}
\newcommand{\eps}{\mbox{$\epsilon$}}
\newcommand{\Lpi}{\mbox{$\Lambda^2_\pi$}}
\newcommand{\Lrho}{\mbox{$\Lambda^2_\rho$}}
\newcommand{\gevsq}{\mbox{GeV$^2$}}

\DeclareGraphicsExtensions{ .eps , .eps.gz ,
                            .ps  , .ps.gz }

\begin{document}

\title{Charged pion form factor between $\bf Q^2=0.60$ and 2.45 GeV$\bf
^2$. II.\\
Determination of, and results for, the pion form factor}

\author{G.M. Huber}
\affiliation{University of Regina, Regina, Saskatchewan S4S 0A2, Canada}
\author{H.P. Blok}
\affiliation{VU university, NL-1081 HV Amsterdam, The Netherlands}
\affiliation{NIKHEF, Postbus 41882, NL-1009 DB Amsterdam, The Netherlands}
\author{T. Horn}
\affiliation{University of Maryland, College Park, Maryland 20742}
\affiliation{Physics Division, TJNAF, Newport News, Virginia 23606}
\author{E.J. Beise}
\affiliation{University of Maryland, College Park, Maryland 20742}
\author{D. Gaskell}
\affiliation{Physics Division, TJNAF, Newport News, Virginia 23606}
\author{D.J. Mack}
\affiliation{Physics Division, TJNAF, Newport News, Virginia 23606}
\author{V. Tadevosyan}
\affiliation{Yerevan Physics Institute, 375036 Yerevan, Armenia}
\author{J. Volmer}
\affiliation{VU university, NL-1081 HV Amsterdam, The Netherlands}
\affiliation{DESY, Hamburg, Germany}
\author{D. Abbott}
\affiliation{Physics Division, TJNAF, Newport News, Virginia 23606}
\author{K. Aniol}
\affiliation{California State University Los Angeles, Los Angeles, California
  90032}
\author{H. Anklin}
\affiliation{Florida International University, Miami, Florida 33119}
\affiliation{Physics Division, TJNAF, Newport News, Virginia 23606}
\author{C. Armstrong}
\affiliation{College of William and Mary, Williamsburg, Virginia 23187}
\author{J. Arrington}
\affiliation{Physics Division, Argonne National Laboratory, Argonne, Illinois
  60439}
\author{K. Assamagan}
\affiliation{Hampton University, Hampton, Virginia 23668}
\author{S. Avery}
\affiliation{Hampton University, Hampton, Virginia 23668}
\author{O.K. Baker}
\affiliation{Hampton University, Hampton, Virginia 23668}
\affiliation{Physics Division, TJNAF, Newport News, Virginia 23606}
\author{B. Barrett}
\affiliation{Saint Mary's University, Halifax, Nova Scotia, Canada}
\author{C. Bochna}
\affiliation{University of Illinois, Champaign, Illinois 61801}
\author{W. Boeglin}
\affiliation{Florida International University, Miami, Florida 33119}
\author{E.J. Brash}
\affiliation{University of Regina, Regina, Saskatchewan S4S 0A2, Canada}
\author{H. Breuer}
\affiliation{University of Maryland, College Park, Maryland 20742}
\author{C.C. Chang}
\affiliation{University of Maryland, College Park, Maryland 20742}
\author{N. Chant}
\affiliation{University of Maryland, College Park, Maryland 20742}
\author{M.E. Christy}
\affiliation{Hampton University, Hampton, Virginia 23668}
\author{J. Dunne}
\affiliation{Physics Division, TJNAF, Newport News, Virginia 23606}
\author{T. Eden}
\affiliation{Physics Division, TJNAF, Newport News, Virginia 23606}
\affiliation{Norfolk State University, Norfolk, Virginia}
\author{R. Ent}
\affiliation{Physics Division, TJNAF, Newport News, Virginia 23606}
\author{H. Fenker}
\affiliation{Physics Division, TJNAF, Newport News, Virginia 23606}
\author{E.F. Gibson}
\affiliation{California State University, Sacramento, California 95819}
\author{R. Gilman}
\affiliation{Rutgers University, Piscataway, New Jersey 08855}
\affiliation{Physics Division, TJNAF, Newport News, Virginia 23606}
\author{K. Gustafsson}
\affiliation{University of Maryland, College Park, Maryland 20742}
\author{W. Hinton}
\affiliation{Hampton University, Hampton, Virginia 23668}
\author{R.J. Holt}
\affiliation{Physics Division, Argonne National Laboratory, Argonne, Illinois
  60439}
\author{H. Jackson}
\affiliation{Physics Division, Argonne National Laboratory, Argonne, Illinois
  60439}
\author{S. Jin}
\affiliation{Kyungpook National University, Taegu, Korea}
\author{M.K. Jones}
\affiliation{College of William and Mary, Williamsburg, Virginia 23187}
\author{C.E. Keppel}
\affiliation{Hampton University, Hampton, Virginia 23668}
\affiliation{Physics Division, TJNAF, Newport News, Virginia 23606}
\author{P.H. Kim}
\affiliation{Kyungpook National University, Taegu, Korea}
\author{W. Kim}
\affiliation{Kyungpook National University, Taegu, Korea}
\author{P.M. King}
\affiliation{University of Maryland, College Park, Maryland 20742}
\author{A. Klein}
\affiliation{Old Dominion University, Norfolk, Virginia 23529}
\author{D. Koltenuk}
\affiliation{University of Pennsylvania, Philadelphia, Pennsylvania 19104}
\author{V. Kovaltchouk}
\affiliation{University of Regina, Regina, Saskatchewan S4S 0A2, Canada}
\author{M. Liang}
\affiliation{Physics Division, TJNAF, Newport News, Virginia 23606}
\author{J. Liu}
\affiliation{University of Maryland, College Park, Maryland 20742}
\author{G.J. Lolos}
\affiliation{University of Regina, Regina, Saskatchewan S4S 0A2, Canada}
\author{A. Lung}
\affiliation{Physics Division, TJNAF, Newport News, Virginia 23606}
\author{D.J. Margaziotis}
\affiliation{California State University Los Angeles, Los Angeles, California
  90032}
\author{P. Markowitz}
\affiliation{Florida International University, Miami, Florida 33119}
\author{A. Matsumura}
\affiliation{Tohoku University, Sendai, Japan}
\author{D. McKee}
\affiliation{New Mexico State University, Las Cruces, New Mexico 88003-8001}
\author{D. Meekins}
\affiliation{Physics Division, TJNAF, Newport News, Virginia 23606}
\author{J. Mitchell}
\affiliation{Physics Division, TJNAF, Newport News, Virginia 23606}
\author{T. Miyoshi}
\affiliation{Tohoku University, Sendai, Japan}
\author{H. Mkrtchyan}
\affiliation{Yerevan Physics Institute, 375036 Yerevan, Armenia}
\author{B. Mueller}
\affiliation{Physics Division, Argonne National Laboratory, Argonne, Illinois
  60439}
\author{G. Niculescu}
\affiliation{James Madison University, Harrisonburg, Virginia 22807}
\author{I. Niculescu}
\affiliation{James Madison University, Harrisonburg, Virginia 22807}
\author{Y. Okayasu}
\affiliation{Tohoku University, Sendai, Japan}
\author{L. Pentchev}
\affiliation{College of William and Mary, Williamsburg, Virginia 23187}
\author{C. Perdrisat}
\affiliation{College of William and Mary, Williamsburg, Virginia 23187}
\author{D. Pitz}
\affiliation{DAPNIA/SPhN, CEA/Saclay, F-91191 Gif-sur-Yvette, France}
\author{D. Potterveld}
\affiliation{Physics Division, Argonne National Laboratory, Argonne, Illinois
  60439}
\author{V. Punjabi}
\affiliation{Norfolk State University, Norfolk, Virginia}
\author{L.M. Qin}
\affiliation{Old Dominion University, Norfolk, Virginia 23529}
\author{P.E. Reimer}
\affiliation{Physics Division, Argonne National Laboratory, Argonne, Illinois
  60439}
\author{J. Reinhold}
\affiliation{Florida International University, Miami, Florida 33119}
\author{J. Roche}
\affiliation{Physics Division, TJNAF, Newport News, Virginia 23606}
\author{P.G. Roos}
\affiliation{University of Maryland, College Park, Maryland 20742}
\author{A. Sarty}
\affiliation{Saint Mary's University, Halifax, Nova Scotia, Canada}
\author{I.K. Shin}
\affiliation{Kyungpook National University, Taegu, Korea}
\author{G.R. Smith}
\affiliation{Physics Division, TJNAF, Newport News, Virginia 23606}
\author{S. Stepanyan}
\affiliation{Yerevan Physics Institute, 375036 Yerevan, Armenia}
\author{L.G. Tang}
\affiliation{Hampton University, Hampton, Virginia 23668}
\affiliation{Physics Division, TJNAF, Newport News, Virginia 23606}
\author{V. Tvaskis}
\affiliation{VU university, NL-1081 HV Amsterdam, The Netherlands}
\affiliation{NIKHEF, Postbus 41882, NL-1009 DB Amsterdam, The Netherlands}
\author{R.L.J. van der Meer}
\affiliation{University of Regina, Regina, Saskatchewan S4S 0A2, Canada}
\author{K. Vansyoc}
\affiliation{Old Dominion University, Norfolk, Virginia 23529}
\author{D. Van Westrum}
\affiliation{University of Colorado, Boulder, Colorado 76543}
\author{S. Vidakovic}
\affiliation{University of Regina, Regina, Saskatchewan S4S 0A2, Canada}
\author{W. Vulcan}
\affiliation{Physics Division, TJNAF, Newport News, Virginia 23606}
\author{G. Warren}
\affiliation{Physics Division, TJNAF, Newport News, Virginia 23606}
\author{S.A. Wood}
\affiliation{Physics Division, TJNAF, Newport News, Virginia 23606}
\author{C. Xu}
\affiliation{University of Regina, Regina, Saskatchewan S4S 0A2, Canada}
\author{C. Yan}
\affiliation{Physics Division, TJNAF, Newport News, Virginia 23606}
\author{W.-X. Zhao}
\affiliation{M.I.T.--Laboratory for Nuclear Sciences and Department of Physics, 
      Cambridge, Massachusetts 02139}
\author{X. Zheng}
\affiliation{Physics Division, Argonne National Laboratory, Argonne, Illinois
  60439}
\author{B. Zihlmann}
\affiliation{University of Virginia, Charlottesville, Virginia 22901}
\affiliation{Physics Division, TJNAF, Newport News, Virginia 23606}
\collaboration{The Jefferson Lab \fpi\ Collaboration}
\noaffiliation

\date{\today}
 
\begin{abstract}
The charged pion form factor, \fpi(\qsq), is an important quantity which can be
used to
advance our knowledge of hadronic structure.  However, the extraction of \fpi\ 
from data requires a model of the $^1$H$(e,e^{\prime}\pi^+)n$ reaction, and thus is
inherently model dependent.  Therefore, a detailed description of the
extraction of the charged pion form factor from electroproduction data obtained
recently at Jefferson Lab is presented, with particular focus given to the
dominant uncertainties in this procedure.  Results for \fpi\ are presented for
\qsq=0.60-2.45 \gevsq.  Above $Q^2=1.5$ \gevsq, the \fpi\ values are
systematically below the monopole parameterization that describes the low \qsq\
data used to determine the pion charge radius.  The pion form factor can be 
calculated in a wide variety of theoretical approaches, and the experimental
results are compared to a number of calculations.  This comparison is
helpful in understanding the role of soft versus hard contributions to hadronic
structure in the intermediate \qsq\ regime.
\end{abstract}


\maketitle
 
\section{Introduction}

There is much interest in trying to understand the structure of hadrons, both
mesons and baryons, in terms of their constituents, the quarks and gluons.
However, this structure is too complicated to be calculated rigorously in
Quantum Chromodynamics (QCD) because perturbative QCD (pQCD) methods are not
applicable in the confinement regime.  Chiral Perturbation Theory can give
valuable insights, but it is limited to small values of the photon virtuality 
\qsq.  Hence, in the intermediate \qsq\ regime one has to resort to models like
the constituent quark model or methods employing Light-Cone (LC) dynamics or
the Bethe-Salpeter (plus Dyson-Schwinger) equation, or to other approaches 
such as the use of dispersion relations or (QCD or LC) sum rules.

Transitions and (transition) form factors are crucial elements for gauging the
ideas underlying these QCD-based models.  For example, the constituent quark
model gives a fairly good description of the meson and baryon spectrum and some
transitions, but quark effective form factors are typically required when
describing hadronic form factors in the experimentally accessible \qsq\ region.
In this framework, the study of hadronic form factors can thus be viewed as a
study of the transition from constituent to current quark degrees of freedom.
As exemplified by the many calculations of it, the electric form factor of the
pion, \fpi, is one of the best observables for the investigation of the
transition of QCD effective degrees of freedom in the soft regime, governed by
all kinds of quark-gluon correlations at low \qsq, to the perturbative
(including next-to-leading order and transverse corrections) regime at higher
\qsq.

In contrast to the nucleon, the asymptotic normalization of the pion
wave function is known from pion decay.  The hard part of the $\pi^+$ form
factor can be calculated within the framework of pQCD as the
sum of logarithms and powers of \qsq\
\cite{lep79}
\begin{widetext}
\begin{equation}
F_{\pi}(Q^2)=\frac{4\pi C_F\alpha_s(Q^2)}{Q^2}
\Biggl|\Sigma^{\infty}_{n=0}a_n
\Biggl(log(\frac{Q^2}{\Lambda^2})\Biggr)^{-\gamma_n}\Biggr|^2
[1+O(\alpha_s(Q^2),m/Q^2)],
\label{eqn:pqcd}
\end{equation}
\end{widetext}
which in the $Q^2\rightarrow\infty$ limit becomes \cite{lep79,farrar}
\begin{equation}
F_{\pi}(Q^2) \overrightarrow{_{Q^2 \rightarrow \infty}} 
\frac{16 \pi \alpha_s(Q^2) f_{\pi}^2}{Q^2}, 
\label{eqn:asymp}
\end{equation}
where $f_{\pi}=93$ MeV is the pion decay constant \cite{dumbrajs}. 

Because the pion's $\bar{q}q$ valence
structure is relatively simple, the transition from ``soft''
(non-perturbative) to ``hard'' (perturbative) QCD is expected to occur at
significantly lower values of \qsq\ for \fpi\ than for the nucleon form factors
\cite{isg84}.  Some estimates \cite{braun} suggest that pQCD contributions
to the pion form factor are already significant at $Q^2\geq 5$ \gevsq.
On the other hand, a recent analysis \cite{brag07} indicates that
non-perturbative contributions dominate the pion form factor up to relatively
large values of \qsq, giving more than half of the pion form factor up to
\qsq=20 \gevsq.  Thus, there is an ongoing theoretical debate on the 
interplay of these hard and soft components at intermediate \qsq, and high
quality experimental data are needed to help guide this discussion.

In this work, we concentrate exclusively on the spacelike region of the pion
form factor.  For recent measurements in the timelike region see
Ref. \cite{cleofpi}.  At low values of \qsq, where it is governed by
the charge radius of the pion, \fpi\ has been determined up to
\qsq=0.253 \gevsq~\cite{dal82,ame86} from the scattering
of high-energy pions by atomic electrons.
For the determination of the pion form factor at higher
values of \qsq, one has to use high-energy electroproduction of pions on a
nucleon, i.e., employ the $^1$H$(e,e^{\prime}\pi^+)n$ reaction.  For selected
kinematic conditions, the longitudinal cross section is very
sensitive to the pion form factor.  In this way, data for \fpi\ have been
obtained for values of \qsq\ up to 10 \gevsq\ at
Cornell~\cite{beb74,beb76,beb78}.
However, those data suffer from relatively large statistical and systematic
uncertainties.  More precise data were obtained at the Deutsches
Elektronen-Synchrotron (DESY) \cite{ack78,bra79}.
With the availability of high-intensity electron beams, combined with accurate
magnetic spectrometers at the Thomas Jefferson National Accelerator Facility
(JLab), it has been possible to determine L/T separated cross sections with
high precision.  The measurement of these cross sections in 
the regime of \qsq=0.60-1.60 \gevsq\ [Experiment Fpi-1 \cite{vol01,tad07}] and
\qsq=1.60-2.45 \gevsq\ [Experiment Fpi-2 \cite{hor06}] are described in detail
in the preceding paper \cite{paper1}.  
In this paper, it is discussed how to
determine \fpi\ from measured longitudinal cross sections, the values
determined from the JLab and DESY data are presented, and the results of
various theoretical calculations are compared with the experimental data.

Since the pion in the proton is virtual (off its mass-shell), the extraction of
\fpi\ from the measured electroproduction cross sections requires some model or
procedure.  In the next section, the methods that have been used to determine
\fpi\ from the data are discussed.  Section \ref{sec:fpi_vgl} presents the
adopted extraction method and the values of \fpi\ thus determined, including a
full discussion of the uncertainties resulting from the experimental data and
those from the adopted extraction procedure.  Various model calculations of
\fpi\ are discussed and compared to the data in section \ref{sec:models}.  In
the final section, some conclusions are drawn and an outlook for the future is
given.

\section{Methods of determining the pion charge form factor from data
\label{sec:eepi_models}}

The measurement of the pion form factor is challenging.  As stated
in the introduction, at low \qsq\ \fpi\ can be measured in a model-independent
manner via the elastic scattering of $\pi^+$ from atomic electrons, such as has
been done up to \qsq=0.253 \gevsq\ at Fermilab \cite{dal82} and at the CERN SPS
\cite{ame86}.  It is not possible to access significantly higher values of
\qsq\ with this technique because of limitations in the energy of the pion
beam together with the unfavorable momentum transfer.
Therefore, at higher values of \qsq\ \fpi\ must be determined from pion
electroproduction on the proton.  The dependence on \fpi\ enters the
cross section via the $t$-channel process, in which the incident electron
scatters from a virtual pion, bringing it on-shell.  This process dominates
near the pion-pole at $t=m_{\pi}^2$, where $t$ is the Mandelstam variable.  The
physical region for $t$ in pion electroproduction is negative, so measurements
should be performed at the smallest attainable values of $-t$.  To minimize
background contributions, it is also necessary to separate out the longitudinal
cross section \sigl, via a Rosenbluth L/T(/LT/TT) separation \cite{sul70}.  

The minimum physical value of $-t$, $-t_{min}$, is non-zero and increases with
increasing \qsq\ and decreasing value of the invariant mass, $W$, of the produced
pion-nucleon system.
Carlson and Milana \cite{carlson} have estimated an approximate upper limit
for the value of $-t_{min}$ of the data appropriate for
the extraction of the pion form factor by studying the competing
non-pole QCD processes, which may complicate the extraction of \fpi\ at higher
\qsq. They found that the background ratio $M_{pQCD}/M_{pole}$ rises
dramatically once $-t_{min}>0.20$ \gevsq.  Their concern stemmed 
from the large value of $-t$ in some of the Cornell results, which have
$-t_{min}> 0.4$ \gevsq~\cite{beb78}.  Therefore, reliable
\fpi\ measurements should be performed at smaller $-t$ and thus higher $W$ (for a
fixed \qsq).  The results presented in this paper respect this $-t_{min}<0.20$
\gevsq\ upper limit.  It is yet to be determined if reliable \fpi\ measurements
can be made in the future at larger $-t$.

The value of $F_{\pi}(Q^2)$ can then be determined from the data by trying to
extrapolate the measured longitudinal cross sections at small values of $-t$ to
the pole at $t=m_{\pi}^2=0.02$ \gevsq, or by comparing the measured longitudinal cross
section at small values of $-t$ to the best available model for the
$^1$H$(e,e^{\prime}\pi^+)n$ reaction, adjusting the value of \fpi\ in the latter.  The
presence of the nucleon and its structure complicates the theoretical model
used, and so an unavoidable implication of this method is that the extracted
pion form factor values are model dependent.  The differential cross sections
\sigl\ versus $t$ over some range of \qsq\ and $W$ are the actual observables
measured by the experiment.  It is important to note that in all cases the use
of a model to extract \fpi\ is justified only if the model correctly predicts
the $t$-dependence and magnitude of the
\sigl\ data as well as the dependence on the invariant mass $W$ of the
photon-nucleon system.

\subsection{Chew-Low Extrapolation Method}

Frazer \cite{frazer} originally proposed that \fpi\ be extracted from \sigl\
via a kinematic extrapolation to the pion-pole, and that this be done in an
analytical manner using the so called Chew-Low extrapolation \cite{chew}.  The
used Born formula is not gauge invariant \cite{van97}, but in principle should
give \fpi, nonetheless, when extrapolating to the pole.

The last serious attempt to extract the space-like pion form factor from
electroproduction data via the Chew-Low method was by Devenish and Lyth
\cite{dev72} in 1972.  Most of the data used were unseparated cross sections.
The extrapolation failed to produce a reliable result, because different
polynomial fits that were equally likely in the physical region gave divergent
values of the form factor when extrapolated to the pion-pole at $t=m_{\pi}^2$.
Since then, the quality of the $\pi^+$ electroproduction data-set has improved
immensely, and separated longitudinal cross sections can now be used, avoiding
the complications stemming from the other parts of the cross section.
Therefore, it has been suggested to us that it may be appropriate to revisit the 
Chew-Low extrapolation method.

However, before trying this method on the new data, it should be tested to see
how reliably one can extrapolate to the pole.  We start with high precision 
\sigl\ `pseudodata' generated as a function of $-t$ with the VGL Regge model.  
This model gives 
a fair to good description of a wide body of pion photo- and
electroproduction data (see section~\ref{sec:vgl}).  The kinematic conditions
for the test are $Q^2=1.594$ \gevsq\ and $W=2.213$ GeV, similar to our
Fpi-2 data.  The input value of the pion form factor in the model
was $F_{\pi}=0.244$.  The model \sigl\ cross sections were then used in a
Chew-Low type extrapolation, with the challenge being to see if the Chew-Low
extrapolation is able to reproduce (within fitting uncertainties) the input
\fpi-value.

The basis of the Chew-Low method is the Born-term model (BTM) formula for the
pion-pole contribution to \sigl.  We use the BTM of Actor, Korner and Bender
\cite{act74}, where pion-pole contribution to \sigl\ is given by
\begin{equation}
N \frac{d\sigma_L}{dt} = 4 \hbar c\ (e g_{\pi NN})^2\ 
\frac{-t\ }{(t-m_{\pi}^2)^2}\ Q^2 F_{\pi}^2(Q^2),
\label{eqn:chewlow}
\end{equation}
where $e^2/(4\pi\hbar c)=1/137$ and $N$, which depends on the flux factor
used in the definition of $\frac{d\sigma_L}{dt}$, is given in our case by
\begin{equation}
N=32\pi (W^2-m_p^2)\sqrt{(W^2-m_p^2)^2+Q^4+2Q^2(W^2+m_p^2)}
\end{equation}
\cite{hand63,van07}.  A monopole parameterization of the
$g_{\pi NN}$ form factor is typically used to determine its value at $t$-values
away from the pion-pole
\begin{equation}
g_{\pi NN}(t)=g_{\pi NN}(m_{\pi}^2)
\Bigl(\frac{\Lambda_{\pi}^2-m_{\pi}^2}{\Lambda_{\pi}^2-t}\Bigr),
\end{equation}
where $g_{\pi NN}(m_{\pi}^2)$ is the experimental value of 13.4 \cite{koch}.
This is also the value used in the VGL calculations.  We use the
$\Lambda_{\pi}=0.80$ GeV result from the QCD Sum Rules calculation by
T. Meissner \cite{mei95}, but because of the extrapolation to the pole the
final result does not depend significantly upon the value chosen.

\begin{figure}[t!]
\begin{center}
\includegraphics[width=8.5cm]{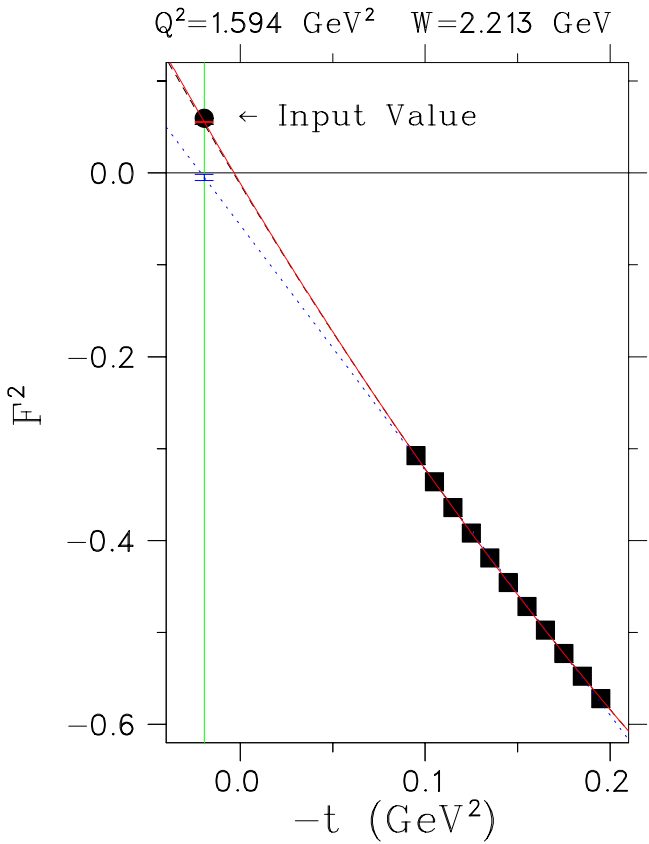}
\caption{(Color online)
  Linear (dotted), quadratic (dashed) and cubic (solid line) extrapolations of
  $F^2$ to the pole as computed from Eqn. \protect{\ref{eqn:chewlow2}}.  
  The boxes are a VGL Regge model calculation for \sigl\ at
  fixed $W=2.213$ GeV and $Q^2=1.594$ \gevsq, calculated with
  \fpi=0.244.  The lower limit of the box range is the
  kinematic endpoint of these \qsq, $W$ values, while the upper limit is given
  by the $t$-range of our experiment.  The input \fpi\ value in the model is
  indicated by the bullet placed at the pion-pole.
\label{fig:chewlow}}
\end{center}
\end{figure}

For the Chew-Low extrapolation, one plots the value of
\begin{equation}
F^2=\frac{N}{4\hbar c\ (e g_{\pi NN})^2}
\frac{(t-m_{\pi}^2)^2}{-Q^2 m_{\pi}^2}\frac{d\sigma_L}{dt}
\label{eqn:chewlow2}
\end{equation}
versus $-t$, which for a pure pole cross section gives a straight line passing
through the origin, with value $F_{\pi}(Q^2)$ at the pole ($t=m_{\pi}^2$).
Other contributions to the cross section, which have to be present,
because the pole contribution alone is not gauge invariant, will change this
behavior, but since they do not contain the $\frac{1}{(t-m_{\pi}^2)^2}$
factor, they will not influence the value of $F^2$ at the pole.  However, it is
not a priori given that the behavior as function of $-t$ is linear, quadratic, or
of higher order, thus introducing a `model' (extrapolation) uncertainty.

Values of $F^2$ for the generated pseudodata, together with linear, quadratic
and cubic extrapolations to the pole are shown in Fig.~ \ref{fig:chewlow}.
Also shown is the input form factor value in the VGL model, plotted at the
pion-pole.  
Quadratic and higher-order extrapolations are almost indistinguishable and give
a very good description of the (pseudo)data, but miss the input value of \fpi.
This was true for all cases that were investigated, from $Q^2=0.60$ to 2.45 \gevsq,
the deviation from the input \fpi-value being 6-15\%, depending on the case
and the order of the extrapolation polynomial.  Overall, there was no
consistent trend for the order of polynomial which was best able to reproduce
the input form factor value.

This study indicates that even if \sigl\ is very well known over a
range of physically-accessible $t$, the
Chew Low extrapolation yields inconsistent results.  
The extrapolated result depends greatly upon the choice of quadratic
cubic, or higher-order function, which all give a very good description of
the data in the physical region.  This indicates that the $t$-dependence of
data in the physical region is insufficient to uniquely constrain the
extrapolation through the unphysical region to the pole, even if the data have
small relative uncertainties.  
Furthermore, even though modern data such as the JLab \sigl\ data are much more
precise than those previously available, they still comprise 4-6 $t$-bins only,
each with statistical and systematic uncertainties of 5-10\%.
Therefore, any polynomial extrapolation of such data to the pole will be 
more unreliable than the pseudodata test case shown here.
Therefore, the Chew-Low extrapolation technique cannot
be used to reliably determine the pion form factor from a realistic \sigl\
data set.

\subsection{Early Extractions of $\bf F_{\pi}$
\label{sec:early}}

Brown {\em et al.} \cite{bro73} at CEA were the first to embrace the use of
theoretical input to determine \fpi\ from their data.  They used the model
of Berends \cite{berends}, which includes the dominant isovector Born
term, with corrections for $t$ values away from the pole by means of fixed-$t$
dispersion relations.  This model was also used by Bebek {\em et al.} for the
analysis of the first two sets of Cornell data~\cite{beb74,beb76}.  The model
gave a fair description of the data, but systematically underpredicted the LT
term of the cross section and the $t$-dependence of the data.

Until then, data were obtained at one (larger) value of the photon polarization
parameter $\epsilon$ only. In the third Cornell experiment~\cite{beb78}, data
were taken at low values of $\epsilon$, so that in combination with the earlier
data an L/T separation could be performed at \qsq-values of 1.19, 2.00 and
3.32~\gevsq.  The value of \sigt\ was found to be substantially larger than
predicted by Berends, especially at larger \qsq.  The values obtained for
\sigl\ had such large error bars that they were not used to determine
\fpi.  Instead, use was made of the observation that within the experimental
error bars the \qsq-dependence of the forward transverse cross section was
satisfactorily reproduced by the \qsq-dependence of the total
virtual-photoproduction cross section.  Therefore, \sigt(\qsq) was
parameterized with the overall scale as a free parameter, and the parameterized
values then used to subtract \sigt\ from the measured unseparated cross
sections to obtain \sigl.  These \sigl\ data at the lowest value of $-t$
were used to determine \fpi, assuming that \sigl\ is given there by the
$t$-channel one pion-exchange Born term.  This was done for all data obtained
at CEA and Cornell.  The uncertainties in \fpi\ thus obtained and presented in
Ref. \cite{beb78} are statistical ones only, and do not include the
contribution from the uncertainty in the value of \sigt\ used in the
subtraction.  Especially at the larger values of \qsq, these are considered to
be substantial, as can be seen from Fig. 4 of Ref.~\cite{beb78}.

The DESY experiments produced high-quality separated cross sections at
$Q^2=0.35$ \gevsq, $W=2.10$ GeV \cite{ack78} and $Q^2=0.70$ \gevsq, $W=2.19$
GeV \cite{bra79}.  Both of these experiments used the generalized Born Term
Model of Gutbrod and Kramer~\cite{gut72} to determine \fpi.  This BTM
incorporates $t$, $s$, and $u$-channel diagrams for the
$\gamma_v+p\rightarrow\pi^+ +n$ reaction, giving a fair description of the
magnitude of the measured unseparated cross sections, but failing to describe
\sigtt\ and \siglt.  However, Gutbrod and Kramer found that when treating the
magnitude of the nucleon form factor $G^p_E(Q^2)$ as a free parameter, a much
better description of the then available data was obtained.  In addition, they
included a factor $e^{t/M^2}$ in order to improve the description of the $t$
dependence of the data.  The justification given is that the nucleon is far
off its mass-shell, whereas the pion is near to its pole.  This generalized BTM
gave a good overall description of the DESY data.  However, at $Q^2=0.70$
\gevsq, nucleon form factors about 50\% above their on-mass-shell values were
needed.  The size of the modification needed at $Q^2=0.35$ \gevsq\ is not
given.

\subsection{Newer Models \label{sec:vgl}}

More recently, two new models for the $^1$H$(e,e^{\prime}\pi^+)n$ reaction have
become available.

In Refs.~\cite{gui97,van97}, Vanderhaeghen, Guidal and Laget (VGL) have
presented a Regge model for pion production in which the pole-like propagators
of Born term models are replaced with Regge propagators, i.e., the interaction
is effectively described by the exchange of a family of particles with the same
quantum numbers instead of a single particle. If the same vertices and coupling
constants are used, the Regge model and the BTM calculations agree at the pole
of the exchanged particle, but away from the pole the Regge model provides a
superior description of the available data.  For forward pion production, the
dominant exchanges are the $\pi$ and $\rho$ trajectories. These determine the
$t$-dependence of the cross section without the use of a $g_{\pi NN}(t)$
factor.  At low values of $-t$, as covered by this work, \sigl\ is
completely determined by the $\pi$~trajectory, while \sigt\ is
also sensitive to the $\rho$ exchange contribution.
Since the $t$-channel $\pi$ diagram is by itself not gauge invariant,
the $s$-channel (for $\pi^+$ production) or $u$-channel (for $\pi^-$
production) nucleon exchange diagram was also Reggeized, to ensure gauge
invariance of their sum.

The VGL model was first applied to pion photoproduction \cite{gui97} and
later extended to electroproduction~\cite{van97}, with monopole forms for the 
$\pi\pi\gamma$ and $\rho\pi\gamma$ form factors: 
\begin{equation}
\label{eqn:monopole}
F_{\pi ,\rho}(Q^2) = [1 + Q^2/\Lambda^2_{\pi ,\rho}]^{-1}.
\end{equation}
Apart from the $\pi\pi\gamma$ and $\rho\pi\gamma$ form factors, the
model is parameter free, as the coupling constants at the vertices (such as
$g_{\rho\pi\gamma}$) are well determined by precise studies and analyses in the
resonance region.  The model gives a good description of the $W$-
and $t$-dependences of then available $\pi^+$ and $\pi^-$ photoproduction data,
including the spin asymmetries, and of the earlier electroproduction data.

\begin{figure*}[h]
\begin{center}
        \includegraphics[height=17.cm, angle=90.]{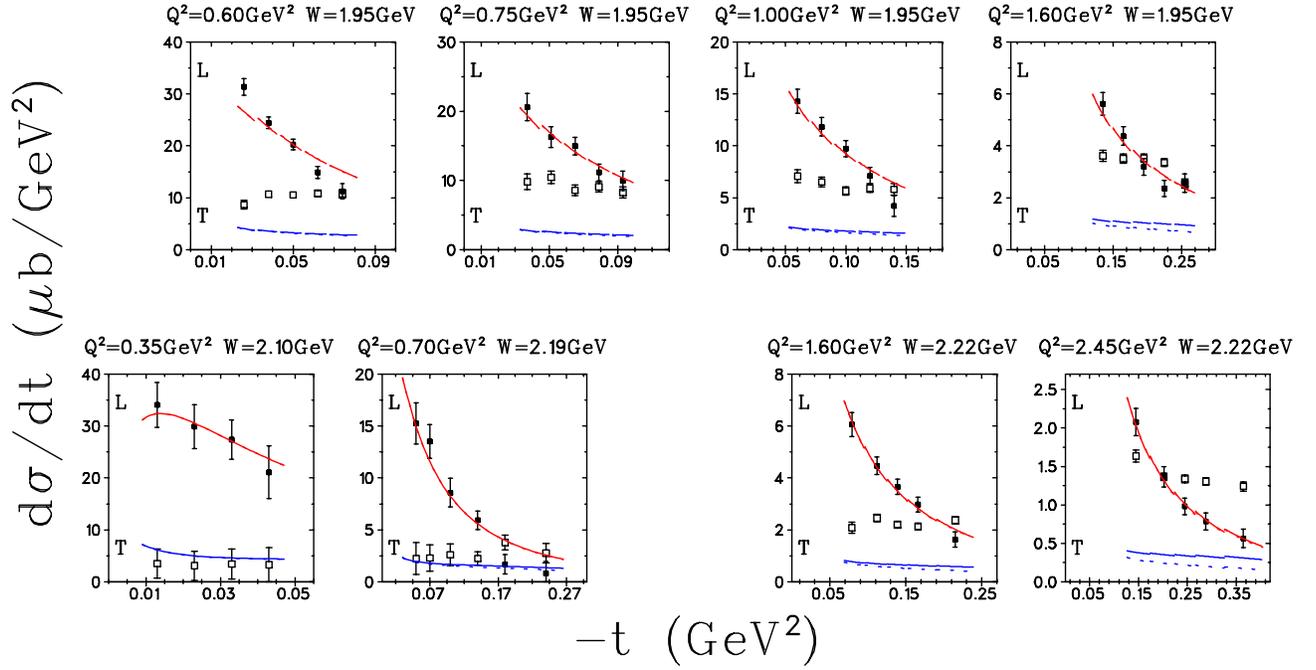}
\end{center}
        \caption{(Color online)
        Separated $\pi^+$ electroproduction cross sections \sigl\ [solid] and
        \sigt\ [open] from JLab and DESY in comparison to the predictions of
        the VGL Regge model \protect{\cite{van97}}.
        The error bars of the JLab data represent the combination of
        statistical and $t$ uncorrelated systematic uncertainties.
        In addition, there is an overall systematic uncertainty of about 6\%,
        mainly from the $t$ correlated, \eps\ uncorrelated systematic
        uncertainty.  The VGL Regge model calculations for \qsq=0.60-1.60
        \gevsq, $W$=1.95 GeV use \Lpi=0.394, 0.372, 0.411, 0.455, 
        \gevsq, and those for \qsq=0.35-2.45 \gevsq, $W\sim 2.1$ GeV use
        \Lpi=0.601, 0.519, 0.513, 0.491 \gevsq.  The solid(dashed) curves
        indicate the \Lrho=1.500(0.600) \gevsq\ value used.
        \label{fig:xsec_vgl}
        }       
\end{figure*}

The VGL predictions have been compared to our measured cross sections
and the ones taken at DESY~\cite{ack78,bra79} in Ref.~\cite{paper1}.
For the discussion in this paper, the data for \sigl\ and \sigt\ are reproduced
in Fig. \ref{fig:xsec_vgl}, together with the results of the model calculations.
The VGL cross sections were evaluated at the same $\overline W$ and
$\overline Q^2$ values as the data, resulting in the discontinuities shown.
The values of \Lpi\ shown are determined by the fitting of the VGL model to the
measured \sigl-values at the five values of $t$ at each \qsq, resulting in
values between 0.37 and 0.51 \gevsq.  The value of \Lrho\ is more poorly known.
Calculations with both \Lrho=0.600 and 1.500 \gevsq\ are shown,
where the upper value is taken from the application of the VGL model to
kaon electroproduction \cite{gui00}.

The model gives an overall good description of our \sigl\ data and those of
\cite{ack78,bra79}, but the description of the $t$-dependence
of the data is worse at \qsq=0.60 and 0.70 \gevsq. 
The poorer description of the \sigl\ data by the VGL model at lower
\qsq\ and $W$ may be due to contributions from resonances, which are not
included explicitly in the Regge model.  This is supported by the fact that the
discrepancy in the $t$-dependence of the \sigl\ data is strongest at the
lowest \qsq\ value, at higher \qsq\ the resonance form factor supposedly
reducing such contributions. 
The values of \sigt\ are severely underestimated, especially at larger \qsq,
even when taking a hard $\rho\pi\gamma$ form factor.
Since the data at the real-photon point are well described, this suggests
that another mechanism, whose contribution increases with \qsq, is at play
\cite{laget}.
Recently the VGL model was extended~\cite{giessen} by including a hard
scattering between the virtual photon and a quark, the latter hadronizing in
combination with the spectator diquark into a pion plus residual nucleon.
With plausible assumptions, a good description of
\sigt\ was obtained, with no influence on \sigl.  Those results support the
idea that the discrepancy in the magnitude of \sigt, which increases with \qsq,
and the discrepancy in the slope of \sigl\ with $-t$, which decreases with
\qsq, are not directly related.  Strategies for dealing with
the latter discrepancy when extracting the pion form factor are discussed in
Sec. \ref{sec:fpi_vgl}.

We also considered a modification to the VGL Regge model published by
J.M. Laget in 2004 \cite{laget04}.  Laget introduces a $t$-dependent factor
into the pion form factor which is related to the pion saturating Regge
trajectory, approaching -1 as $t\rightarrow -\infty$.  The effect of this
modification is to boost \sigt\ by 40\% for the largest $-t$ spanned by our
data (\qsq=2.703 \gevsq, $-t$=0.365 \gevsq), and converging with the unmodified
calculation at small $-t$.  The effect on \sigl\ is under 1\% for the largest
$-t$ covered by our data, and is negligible at $-t_{min}$.

\begin{figure*}[h]
\begin{center}
        \includegraphics[height=17.cm, angle=90.]{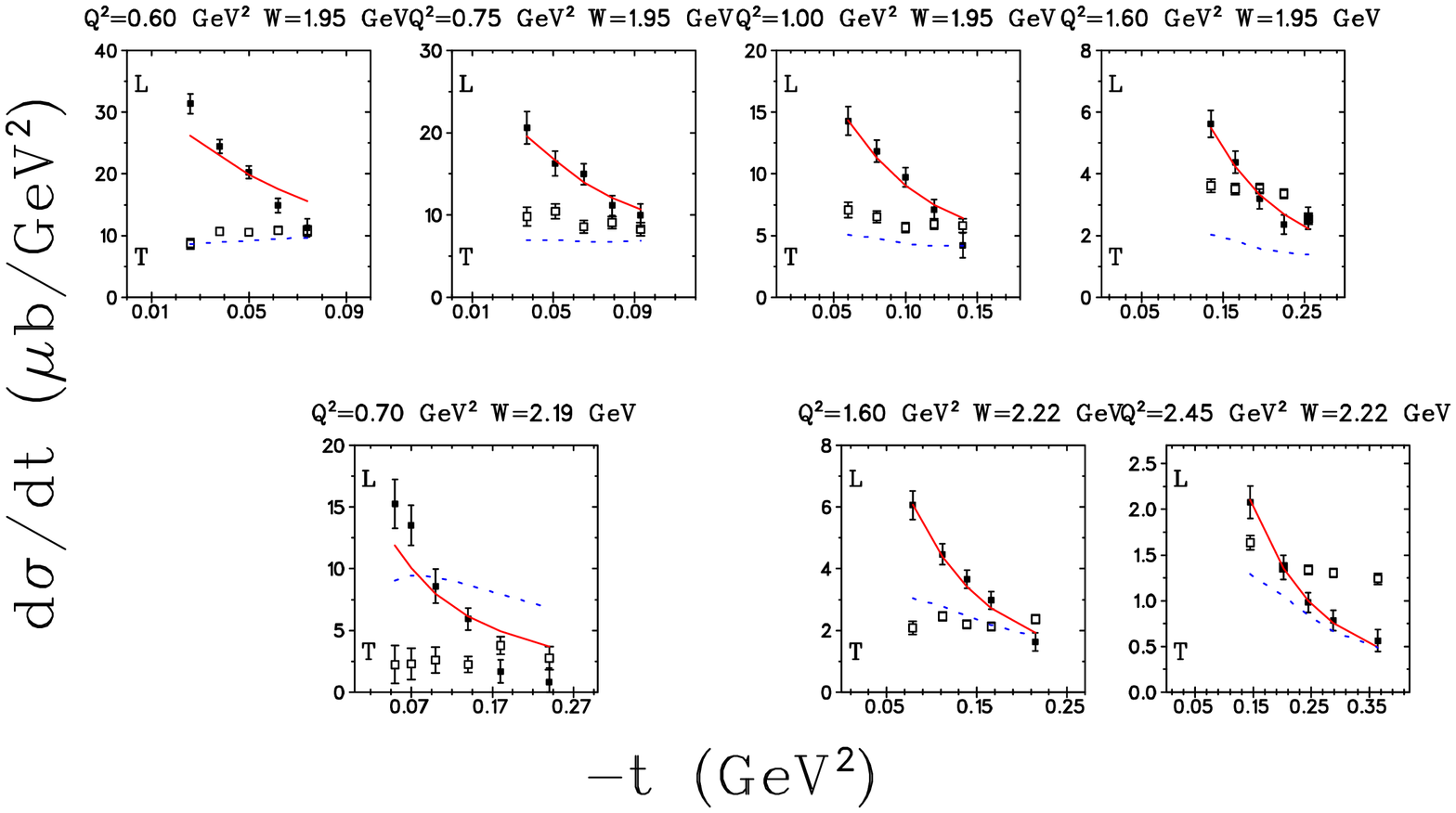}
\end{center}
        \caption{(Color online)
        Separated $\pi^+$ electroproduction cross sections \sigl\ [solid],
        and \sigt\ [open] from this work and DESY \protect{\cite{bra79}} in
        comparison to the FGLO effective Lagrangian model~\protect{\cite{obu07}}.  
        The data error bars and systematic
        uncertainties are as in Fig. \ref{fig:xsec_vgl}. The solid (dashed)
        curves denote model calculations for \sigl\ (\sigt) with \Lpi=0.405,
        0.414, 0.503, 0.654, 0.386, 0.608 and 0.636 \gevsq\ (from upper left to
        lower right).  The calculations were performed at the same
        $\overline W$ and $\overline Q^2$ as the data, with straight lines
        connecting the calculated values.
        \label{fig:xsec_obu}
        }       
\end{figure*}

Another recent development is the effective Lagrangian model of Faessler, 
Gutsche, Lyubovitskij and Obukhovsky (FGLO, Ref.~\cite{obu06,obu07}).  
This is a modified Born Term Model, in which an effective Lagrangian is used to
describe nucleon, pion, $\rho$ and photon degrees of freedom.  The (combined)
effect of $s$- and $u$-channel contributions, which interferes with the pion
$t$-pole, is modeled using a constituent quark model.  The authors show
that the $\rho$ $t$-pole contribution is very important in the
description of the magnitude of \sigt.  When comparing vector and tensor
representations of the $\rho$ contribution, the latter was found to give better
results.  Unlike the VGL model, the \sigl\ cross section depends here also on
the $\rho$ exchange, because of the interference of the $\pi$ and tensor $\rho$
exchange contributions.  The model contains a few free parameters, such as the
renormalization constant of the Kroll-Ruderman contact term used to model the
$s(u)$-channel, and $t$-dependent strong meson-nucleon vertices, which are
parameterized in monopole form, as are the electromagnetic form factors.  The
corresponding parameters were adjusted so as to give overall good agreement
with our \sigl\ and \sigt\ data.

As in case of the VGL model, a detailed comparison of the FGLO model results
to the measured data is given in Ref.~\cite{paper1}, while the results for
\sigl\ and \sigt\ are also shown in Fig. \ref{fig:xsec_obu}.
The values of \Lpi\ used were
determined by the fitting of the model to the \sigl\ $t$-bins at each \qsq,
while keeping the other parameters fixed at the values assigned by the authors.
In some cases, this results in different \Lpi\ values than shown in
Ref. \cite{obu07}.  However, it should be kept in mind that the FGLO model
\sigl\ cross sections also depend on other parameters, which have been adjusted
by the authors of the model to give good agreement to our \sigl\ and \sigt\
data.  To the best of our knowledge, the \qsq=0.7 \gevsq\ data of
Ref.~\cite{bra79} were not taken into account when these parameters were
determined.

Generally, the agreement of the FGLO model with the \sigl\ data is rather good
except for the \qsq=0.60 [Fpi-1] and 0.70 \gevsq\ \cite{bra79} measurements.
There is a serious discrepancy in the $Q^2$- and $W$-dependence of the
\sigt\ data. For \qsq\ around 0.7 \gevsq, the model agrees fairly well with the
data at $W=1.95$ GeV, but it over-predicts the \qsq=0.70 \gevsq, $W=2.19$ GeV
data by a large factor.  On the other hand, for \qsq=1.60 \gevsq, the
$W=1.95$ GeV data are under-predicted by about a factor of two, while those
at $W=2.22$ GeV are reproduced, and the $W=2.22$ GeV data for \qsq=2.45 \gevsq\ 
are under-predicted again by 20-60\%. This indicates some problem in
the description of the $Q^2, W$-dependences of the $\rho$ exchange used to
describe \sigt. Because of the $\rho-\pi$ interference, the problems with the
description of \sigt\ also affect the \sigl\ calculation.  This makes it hard
to estimate how reliable the values of \fpi\ would be if extracted from the
data using this model.

\section{$\bf F_{\pi}$ Results \label{sec:fpi_vgl}}

As already discussed, the separated cross sections versus $t$ over some range
of \qsq\ and $W$ are the actual observables measured by the experiment, and the
extraction of the pion form factor from these data is inherently model
dependent.  Ideally, one would like to have a variety of reliable
electroproduction models to choose from, so that the model dependence of the
extracted \fpi\ values can be better understood.  Since the VGL Regge model is
able, without fitted parameters, to provide a good description of both
$\pi^+$ and $\pi^-$ photoproduction data, and of \sigl\ electroproduction data
over a range in $W$, $t$, and \qsq, it is our opinion that at the moment only
this model has shown itself to be sufficiently reliable to enable its use to
extract pion form factor values from the \sigl\ data.  Therefore, we will use
this model to determine our \fpi\ values.  Clearly, the \fpi\ values determined
are strictly within the context of the VGL Regge model, and other values may
result if other, better models become available in the future.

\subsection{$\bf W\approx 2.2$ GeV Data}

\begin{table}
\begin{center}
\begin{tabular}{cc|c|c}
$Q^2$ & W & \Lpi\ & $F_{\pi}$ \\
(\gevsq) & (GeV) & (\gevsq) & \\ \hline
0.60   & 1.95  & $0.458\pm 0.031^{+0.255}_{-0.068}$ & $0.433\pm 0.017^{+0.137}_{-0.036}$ \\
0.75   & 1.95  & $0.388\pm 0.038^{+0.135}_{-0.053}$ & $0.341\pm 0.022^{+0.078}_{-0.031}$ \\
1.00   & 1.95  & $0.454\pm 0.034^{+0.075}_{-0.040}$ & $0.312\pm 0.016^{+0.035}_{-0.019}$ \\
1.60   & 1.95  & $0.485\pm 0.038^{+0.035}_{-0.027}$ & $0.233\pm 0.014^{+0.013}_{-0.010}$ \\
\hline
0.35   & 2.10  & $0.601\pm 0.060$\ \ \ \ \ \ \ \ \ &
$0.632\pm 0.023$\ \ \ \ \ \ \ \ \ \\
0.70   & 2.19  & $0.627\pm 0.058^{+0.096}_{-0.085}$ & $0.473\pm 0.023^{+0.038}_{-0.034}$ \\
\hline
1.60   & 2.22  & $0.513\pm 0.033^{+0.052}_{-0.022}$ & $0.243\pm0.012^{+0.019}_{-0.008}$ \\
2.45   & 2.22  & $0.491\pm 0.035^{+0.045}_{-0.024}$ & $0.167\pm0.010^{+0.013}_{-0.007}$ \\
\end{tabular}
\end{center}
\caption{\Lpi\ and \fpi\ values from this work, and the reanalyzed data from 
   Refs. \protect{\cite{ack78,bra79}} using the same method. 
   The first error includes all experimental and analysis uncertainties,
   and the second error is the `model uncertainty' as described in the text.
   The total uncertainty is found by taking their sum, in quadrature.
   Please note that in some cases the \Lpi\ value listed is different than the
   value used in Fig. \ref{fig:xsec_vgl}.
\label{tab:fpi}
}
\end{table}

As shown in Fig. \ref{fig:xsec_vgl}, the VGL model does a good job of
describing the $t$-dependence of the \sigl\ cross sections at $W\approx 2.2$ GeV,
$Q^2=$0.35, 1.60 and 2.45 \gevsq.  In these cases, the extraction of the pion form
factor from the data is straightforward: the value of \Lpi\ in the model is
varied until the agreement of the model with the data is optimized.  The mean
$\overline Q^2$ and $\overline W$ values of the data for each $t$-bin are used
when evaluating the model.  \fpi\ is then calculated from
Eqn.~\ref{eqn:monopole}, using the best-fit \Lpi\ and the nominal \qsq\ values.
These are listed in the last two lines of Table \ref{tab:fpi}.

The experimental statistical and systematic uncertainties were propagated to
the \fpi\ uncertainties as follows.  The statistical and
$t,\epsilon$-uncorrelated systematic uncertainties\footnote{These uncertainties 
are described in detail in Ref. \cite{paper1}} were applied to the \sigl\
data prior to the fitting of the VGL model to the \sigl\ data.  This yields the
best-fit \Lpi\ value and its associated fitting uncertainty.  The effects of
the $t$-correlated, $\epsilon$-uncorrelated, and the $t,\epsilon$-correlated
systematic uncertainties on the fit were determined by investigating the
variation in \Lpi\ values allowed by fitting to the lowest $-t$ bin only.  Of
these, the $\epsilon$-uncorrelated, $t$-correlated systematic uncertainty is
amplified by $1/\Delta\epsilon$ in the L-T separation, while the
$t,\epsilon$-correlated uncertainty is not.  The resulting uncertainties are
added in quadrature to the fitting error, yielding the first \Lpi\ uncertainty
listed in Table \ref{tab:fpi}.  This value is also propagated to \fpi\
according to the monopole parameterization, yielding the first \fpi\
uncertainty listed.

\begin{figure}[h]
\begin{center}
        \includegraphics[width=8.5cm]{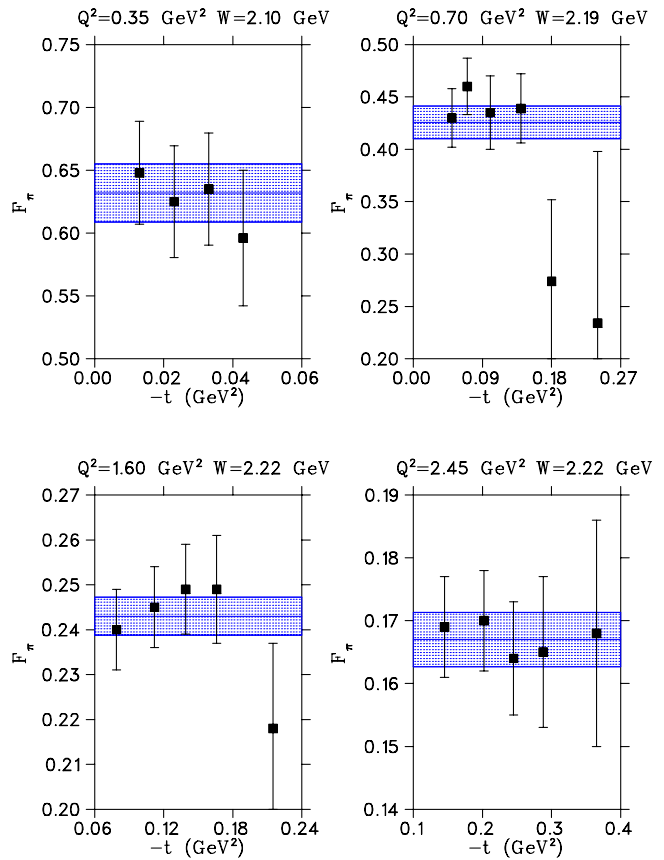}
\end{center}
        \caption{(Color online)
          \fpi\ consistency check for the DESY and Fpi-2 data at $W \approx
          2.2$ GeV.  The solid squares indicate the \fpi\ values that would be 
          obtained if the VGL model was fit to each \sigl\ point separately.
          The shaded band is the \fpi\ value that is obtained if the model is
          fit to all of the $t$-bins.  The error bars and band reflect the
          statistical and $t$-uncorrelated systematic uncertainties only.
        \label{fig:fpi2_check}
        }       
\end{figure}

In order to check if the extracted value of \fpi\ depends on the $t$-range
used, the VGL model (i.e., the value of \Lpi) was fitted separately to each
\sigl\ point from Fpi-2 and DESY \cite{ack78,bra79}, and the
corresponding values of \fpi\ determined.  In order to remove the natural
variation of \fpi\ with the $\overline Q^2$ of each bin, the nominal
\qsq\ values were used in the monopole equation.  A plot of the obtained \fpi\
versus $t$ is shown in Fig. \ref{fig:fpi2_check}.  Also indicated as the shaded
band is the \fpi\ value with the uncertainty that is obtained if one fits to
all of the $t$-bins simultaneously.  Except perhaps at \qsq=0.70 \gevsq, the data
show no residual $t$-dependence beyond the statistical fluctuation.

\subsection{$\bf W=1.95$ GeV Data}

\begin{figure}[h]
\begin{center}
        \includegraphics[width=8.5cm]{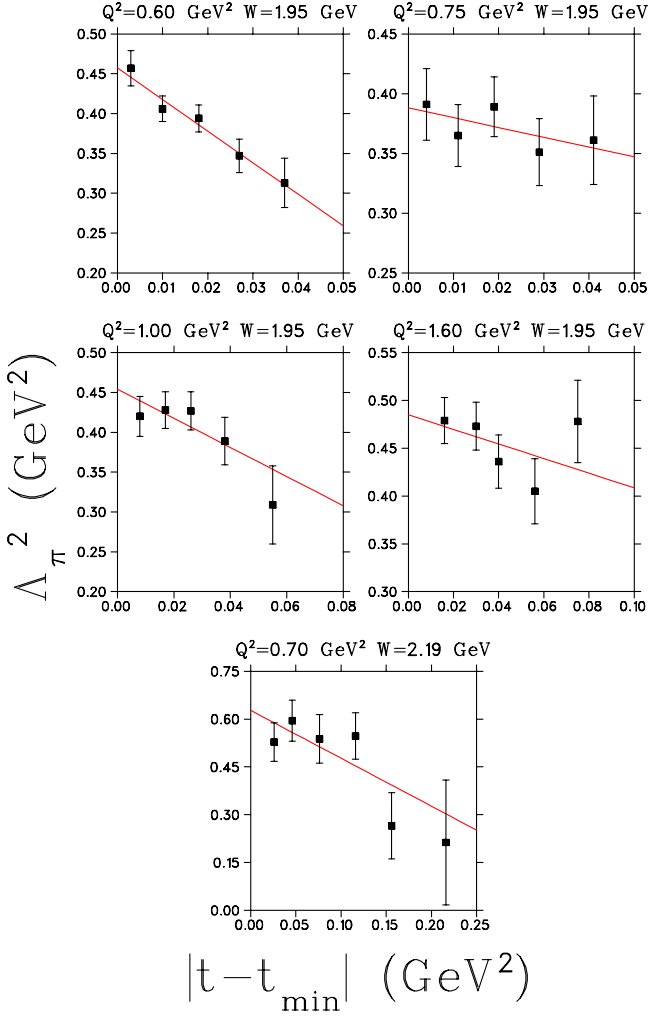}
\end{center}
        \caption{(Color online)
        Values of \Lpi\ determined from the fit of the VGL model to each
        $t$-bin, and linear fit to same.  The error bars reflect the
        statistical and $t$-uncorrelated systematic uncertainties.
        The additional overall systematic uncertainties, which were applied
        after the fit, are not shown.
        \label{fig:fpi1_lpi}
        }       
\end{figure}

As already shown in Sec. \ref{sec:vgl}, the VGL model does not fully describe
the $t$-dependence of our \sigl\ data at $W=1.95$ GeV.  The difficulty, as far
as the \fpi\ extraction is concerned, is that there is no theoretical guidance
for the assumed interfering background not included in the VGL model, even if
one assumes that it is due to resonances.  Virtually nothing is known about the
L/T character of resonances at $W=1.95$ GeV, let alone how they may influence
\sigl\ through their interference with the $\pi$-pole amplitude.  Given this
lack of theoretical guidance, we are forced to make some assumptions in
extracting \fpi\ from these data.  Our guiding principle is to minimize these
assumptions to the greatest extent possible.  The form factor extraction method
that we have adopted for these data relies on the single assumption that the
contribution of the background is smallest at the kinematic endpoint $t_{min}$.

Our best estimate of \fpi\ for the $W=1.95$ GeV data is determined in the
following manner.  Using the value of \Lpi\ as a free parameter, the VGL model
was fitted to each $t$-bin separately, yielding
$\Lambda^2_\pi(\overline{Q^2},\overline{W},t)$ values as shown in
Fig. \ref{fig:fpi1_lpi}.  The values of \Lpi\ tend to decrease as $-t$
increases, presumably because of an interfering background not included in the
VGL model.  Since the pole cross section containing \fpi\ increases strongly
with decreasing $-t$, we assume that the effect of this background will be
smallest at the lowest value of $|t|$ allowed by the experimental kinematics, 
$|t_{min}|$.  Thus, an extrapolation of \Lpi\ to this physical limit is used to
obtain our best estimate of \fpi.  The value of \Lpi\ at $t_{min}$ is obtained
by a linear fit to the data in Fig. \ref{fig:fpi1_lpi}.  The resulting \Lpi\
and \fpi\ values for the Fpi-1 data are listed in Table \ref{tab:fpi}.  The
first uncertainty listed includes both the experimental and the linear fit
extrapolation uncertainties.

Since Fig. \ref{fig:fpi2_check} suggests also a dependence (at larger $-t$)
between the VGL calculation and the \qsq=0.70 \gevsq\ data of
Ref. \cite{bra79}, this \fpi\ extraction method was
also applied to those data.  The result
obtained when extrapolating to $t_{min}$ is listed in Table \ref{tab:fpi}.  The
value of \fpi(\Lpi) is 11(20)\% larger than if the VGL model was simply fit to
all data points.  Applying the same procedure to our $W=2.22$ GeV data, it was
found that the resulting values of \fpi(\Lpi) would be 1(2)\% larger, which is
statistically insignificant, confirming that the $t$-dependence of those data
is well described by the VGL model.

\subsection{Model Uncertainty Estimate}

The fact that we used an additional assumption for the cases where the VGL
model does not completely describe the $t$-dependence of the \sigl\ data
causes an additional uncertainty in the
extracted \fpi\ value, which we term `model uncertainty'.  This model
uncertainty, which is within the context of the VGL model, should be
distinguished from the general model uncertainty discussed in
section~\ref{sec:eepi_models}, which would result when using different models.
In order to make a quantitative estimate of this additional uncertainty, the
spread in extracted values of \Lpi\ (and thus \fpi) was investigated by
assuming specific forms of the interfering background missing in the VGL
model.

An effective upper limit for \fpi\ is obtained by assuming that the
background yields a constant, negative, contribution to \sigl.  For each value
of \qsq, this background and the value of \Lpi\ were fit together to the data,
assuming that the background is constant with $t$.  The fitted contribution of
the background was found to drop strongly with increasing \qsq.  A second
possibility is to assume, besides the VGL amplitude, a $t$-independent
interfering background amplitude, fitting for every \qsq\ the magnitude and
phase of the latter, together with the value of \Lpi.  Although the fitting
uncertainties are very large, the results suggest an interfering amplitude
whose magnitude decreases monotonically with increasing \qsq.  In this case, the
interference between the background amplitude and the VGL amplitude, which
depends on their relative phase, does not necessarily result in a net negative
cross section contribution to \sigl.

The estimated model uncertainty is determined from the spread of the
\Lpi\ values and their uncertainties at each \qsq, obtained with these two
choices of background.  To keep the number of degrees of freedom the same in
both cases, the background was fixed to the value giving the best $\chi^2$, and
\Lpi\ and its uncertainty were then determined in a one-parameter fit of the
VGL model plus background to the data.  Since the statistical uncertainties of
the data are already taken into account in the first given uncertainty in Table
\ref{tab:fpi}, the contribution of the statistical uncertainties of the data
were quadratically removed from the \Lpi\ uncertainties given by the fit.  The
model uncertainties at each \qsq\ are then taken as the range plus corrected
fitting uncertainty given by these two methods, relative to the value of \Lpi\
determined from the extrapolation to $t_{min}$.  This procedure was applied to
all data except those of Ref. \cite{ack78}, yielding the model uncertainties
listed as the second (asymmetric) uncertainty in Table~\ref{tab:fpi}.  No model
uncertainty was calculated for the \qsq=0.35 \gevsq\ data from DESY because the
$t$-range spanned by those data (only 0.03 \gevsq) was too small for this
procedure to be reliably applied.

For the $W=1.95$ GeV data, the model uncertainty in the extracted \fpi\ value
drops from about 20\% at \qsq=0.60 \gevsq\ to about 5\% at 1.60 \gevsq.  To be
consistent, the same procedure was applied to the $W=2.22$ GeV data, which
yielded model uncertainties of about 5\% at both \qsq=1.60 and 2.45 \gevsq.
These rapidly dropping uncertainties with increasing \qsq\ reflect the smaller
discrepancy of the VGL calculation with the $t$-dependence of the data at
larger values of \qsq\ and $W$.  These findings are at least compatible with
the idea that resonance contributions, which presumably have a form factor that
drops rapidly with \qsq, are responsible.  They also suggest that our \fpi\
extraction methods are robust, when the background contribution is small, as
appears to be the case at the higher value of $W$.

\subsection{Discussion and Comparison with Empirical Fits
\label{sec:monopole}}

\begin{figure}[t]
\begin{center}
        \includegraphics[height=8.5cm,angle=90]{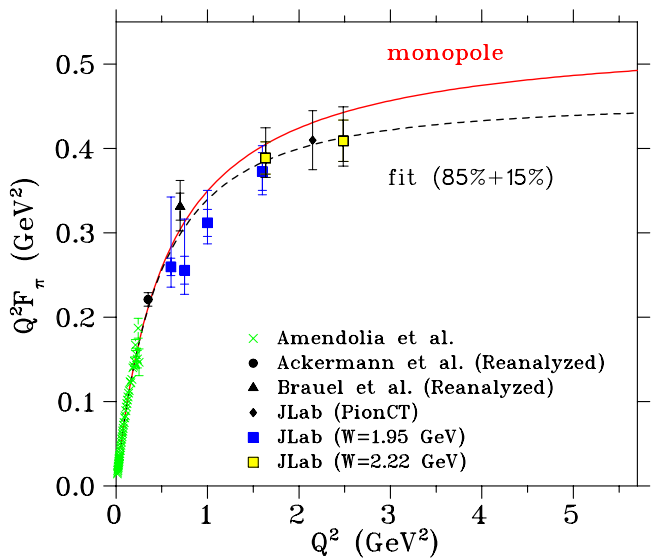}
\end{center}
        \caption{(Color online)
        $Q^2F_{\pi}$ data from this work, compared to previously published data.
        The solid Brauel {\em et al.} \protect{\cite{bra79}} point has been reanalyzed
        as discussed in the text.
        The outer error bars for the JLab data
        and the reanalyzed Brauel {\em et al.} data include all experimental
        and model uncertainties, added in quadrature, while the inner error
        bars reflect the experimental uncertainties only.
        Also shown is the monopole fit by Amendolia {\em et al.}
        \protect{\cite{ame86}} as well as a 85\% monopole+15\% dipole fit
        to our data.
        \label{fig:q2fpi_mono}
        }       
\end{figure}

The form factors extracted from the Fpi-1 and Fpi-2 data with the
use of the VGL model are shown in Fig. \ref{fig:q2fpi_mono}, along with the
reanalyzed \qsq=0.70 \gevsq\ data of Ref. \cite{bra79}, the elastic scattering
measurements of Ref. \cite{ame86}, and the \qsq=0.35 \gevsq\ data of
Ref. \cite{ack78}.  The Cornell data of Refs. \cite{beb74,beb76,beb78} are not
included because, as discussed in section~\ref{sec:early}, they have large
unknown systematic uncertainties.  The excellent agreement between the \qsq=1.6
\gevsq\ form factor values obtained from our $W=1.95$, 2.22 GeV data, despite
their significantly different $t_{min}$ and $W$ values, indicates that the
model uncertainties from the use of the VGL model seem to be under control, at
least in this \qsq-range.  Also shown is a more recently obtained value at
\qsq=2.15 \gevsq\ \cite{pionct}, which was also extracted with the use of
the VGL model.

The solid line shown in Fig. \ref{fig:q2fpi_mono} is the monopole fit obtained
by Amendolia {\em et al.} \cite{ame86} from their elastic scattering data.
This curve is given by
\begin{equation}
F_{mono}=\frac{1}{1+\frac{r_{mono}^2Q^2}{6\hbar^2c^2}},
\end{equation}
where $r_{mono}^2=0.431$ fm$^2$ is their best-fit squared pion charge radius.
Fig. \ref{fig:q2fpi_mono} indicates a systematic departure of the data
from the monopole curve above $Q^2\approx 1.5$ \gevsq.  This departure may have
implications for theoretical approaches that assume the validity of the
monopole parameterization over a wide range of \qsq.

To illustrate the departure from the monopole curve, as well as to
provide an empirical fit that describes the data over the measured \qsq\
range, we also show in Fig \ref{fig:q2fpi_mono} a fit which includes a small
dipole component,
\begin{equation}
F_{fit}=85\%F_{mono}+15\%F_{dip},
\end{equation}
where
\begin{equation}
F_{dip}=\frac{1}{\bigl(1+\frac{r_{dip}^2Q^2}{12\hbar^2c^2}\bigr)^2}
\end{equation}, 
and $r_{dip}^2=0.411$ fm$^2$.  This dipole parameterization has nearly the same
$\chi^2$ for the elastic scattering data as the monopole curve shown
\cite{ame86}, but it drops much more rapidly with \qsq.
The combined monopole plus dipole fit is consistent with our intermediate \qsq\
data, while maintaining the quality of fit to the elastic scattering data.
Since a monopole parameterization does not converge to the pQCD asymptotic
limit (Eqn. \ref{eqn:asymp}), it is expected to fail at some point.  Similarly,
we should expect this empirical monopole+dipole parameterization
to show its limitations when additional high \qsq\ data become available
\cite{fpi12}.

\section{Comparison with Model Calculations
\label{sec:models}}

The pion form factor can be calculated relatively easily in a large number of
theoretical approaches which help advance of our knowledge of
hadronic structure.  In this sense, \fpi\ plays
a role similar to that of the positronium atom in QED.  Here, we compare our
extracted \fpi\ values to a variety of calculations, selected to provide a 
representative sample of the approaches used.

\subsection{Perturbative QCD}

The most firmly grounded approach for the calculation of \fpi\ is that of pQCD.
The large \qsq\ behavior of the pion form factor has already been given in
Eqn. \ref{eqn:pqcd}.  By making use of model-independent dimensional arguments,
the infinitely-large \qsq\ behavior of the pion's quark wave function
(distribution amplitude, or DA) is identified as
\begin{equation}
\phi_{\pi}(x, Q^2\rightarrow\infty)\rightarrow 6f_{\pi}x(1-x)
\end{equation}
whose normalization is fixed from the $\pi^+\rightarrow\mu^+\nu_{\mu}$ decay
constant.  Eqn. \ref{eqn:asymp} follows from this expression.

Neither of these equations is expected to describe the pion form factor in the
kinematic regime of our data, and so much effort has been expended to extend
the calculation of \fpi\ to experimentally accessible \qsq.  In this case, the
pion DA, $\phi_{\pi}(x,\qsq)$, must be determined at
finite \qsq.  Additional effects, such as quark transverse momentum and Sudakov
suppression (essentially a suppression of large quark-quark separation
configurations in elastic scattering processes) must be taken into account.  A
number of authors \cite{melic,stefanis99,stefanis00,ducati,huang04} have
performed leading-twist next-to-leading order (NLO) analyses of \fpi\ at finite
\qsq.  The hard contributions to \fpi\ expand as a leading order part of
order $\alpha_s$ and an NLO part of order $\alpha_s^2$.

\begin{figure}[b]
\begin{center}
        \includegraphics[height=8.5cm,angle=90]{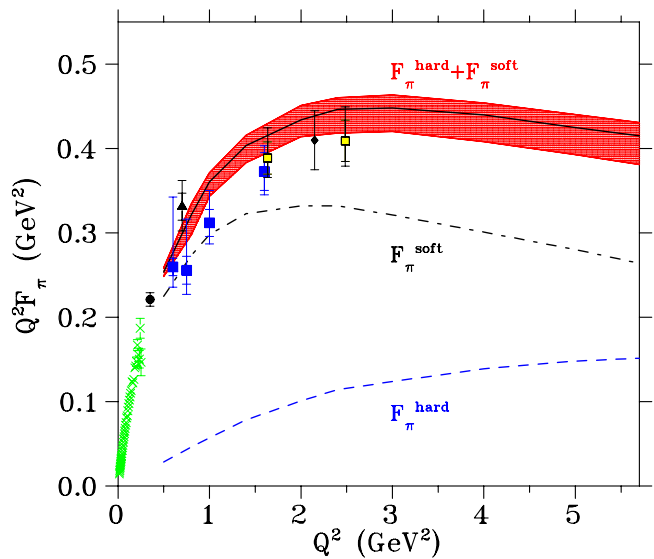}
\end{center}
\caption{(Color online)
The \fpi\ data of Fig. \ref{fig:q2fpi_mono} are compared with a hard LO+NLO
contribution by Bakulev, Passek-Kumericki, Schroers and Stefanis
\protect{\cite{bakulev}} based on an analysis of the pion-photon transition
form factor data from CLEO \protect{\cite{cleo}} and CELLO
\protect{\cite{cello}}.  A soft component, estimated from a local
quark-hadron duality model, is added to bring the calculation into
agreement with the experimental data.  The band around the sum reflects
nonperturbative uncertainties from nonlocal QCD sum rules and renormalization
scheme and scale ambiguities at the NLO level.
\label{fig:pqcd}}
\end{figure}

Bakulev, Passek-Kumericki, Schroers and Stefanis \cite{bakulev} have
investigated the dependence of the form of the DA on the form factors, using
data from a variety of experiments.  These were the $\pi\gamma\gamma$
transition form
factor data from CLEO \cite{cleo} and CELLO \cite{cello}, as well as our \fpi\
data.  Their results are insensitive to the shape of the DA near
$x=1/2$, while its behavior at $x=0$,~1 is decisive.  The resulting hard
contribution to the pion form factor is only slightly larger than that
calculated with the asymptotic DA in all considered schemes.  The result of
their study, shown as $F_{\pi}^{hard}$ in Fig. \ref{fig:pqcd}, is far below our
data.  The drop at low \qsq\ is due to their choice of infrared renormalization,
which is not necessarily shared by other calculations.
To bring the calculation into agreement with the experimental data, a
soft component must also be added.  The treatment of the soft contribution to
the pion DA is model-dependent.  The authors estimate
this soft contribution using a local quark-hadron duality model.  This soft
estimate,
along with the sum of the hard and soft contributions, are also shown in
Fig. \ref{fig:pqcd}.

The interplay at intermediate \qsq\ between the hard and soft components
can be non-trivial, as demonstrated by Braun, Khodjamirian, and Maul
\cite{braun}, using a light-cone sum rule approach.  Their results support a
pion DA that is close to the asymptotic expression, but they find that strong
cancellations between soft terms and hard terms of higher twist lead to the
paradoxical conclusion that the nonperturbative effects in the pion form factor
can be small, and the soft contributions large, simultaneously.  Because of
complications such as these, different theoretical viewpoints on whether the
higher-twist mechanisms dominate \fpi\ until very large momentum transfer, or
not, remain.

\subsection{Lattice QCD}

Unlike QCD-based models, in which confinement must be explicitly added,
Lattice QCD allows calculation from first principles.  However,
while lattice QCD is based on the QCD Lagrangian, it involves a number of
approximations.  Errors are introduced because space and time are crudely
discretized on the lattice.  This error is controlled by the use of improved
lattice QCD actions.  To allow a more rapidly converging action, and hence
reduce CPU usage, the pion mass used is significantly larger than the physical
pion mass.  Chiral extrapolation errors are introduced when the lattice
results, determined with large pion mass, are extrapolated to physical values.
Finally, quenching errors are introduced when disconnected quark loops are
neglected.

\begin{figure}[t]
\begin{center}
\includegraphics[height=8.5cm,angle=90]{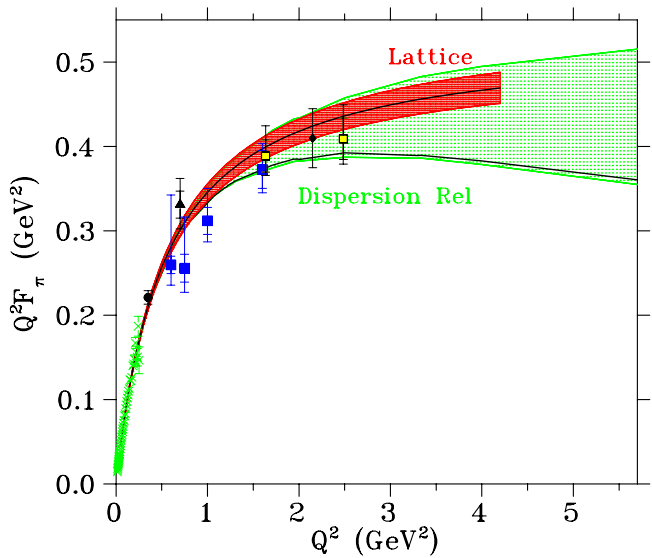}
\end{center}
\caption{(Color online)
The \fpi\ data of Fig. \ref{fig:q2fpi_mono} are compared with the lattice QCD
result of Ref. \protect{\cite{bro07}} and the dispersion relation result of
Ref. \protect{\cite{geshkenbein}}.  The lattice QCD band denotes the
statistical and chiral extrapolation uncertainties in the fit monopole mass to
the simulated data.  The dispersion relation uncertainty band reflects
different assumptions on the distributions of zeroes in the complex $s$-plane,
with the `no zeroes' curve lying close to the `minimum \fpi' limit.
\label{fig:latdisp}}
\end{figure}

The first lattice simulations of \fpi\ were done in the 1980's \cite{woloshyn,
martinelli, draper}.  These pioneering works were primarily a proof of
principle of the lattice technique, and were restricted to \qsq$<1$~\gevsq.
These results are consistent with the low \qsq\ experimental data, within the
large statistical uncertainties of these pioneering calculations.  
Spurred by advances in CPU power and lattice
techniques, as well as the availability of new experimental data, a number of
groups \cite{heide03, nemoto, heide04, rehim, bonnet, boy06, alex06, bro07,
sim07} have returned to the calculation of \fpi\ on the lattice.  Of these, we
compare our data to the recent unquenched simulations of Brommel, {\em et
al.} \cite{bro07}.  They performed simulations for a wide range of pion masses
and lattice spacings, so that both the chiral and continuum limits could be
studied.  However, the lowest pion mass used in the simulations was 400 MeV, so
the chiral extrapolation is significant.  
The authors fitted the \qsq-dependence of each lattice configuration with
a monopole form for the pion form factor and determined the corresponding
monopole mass.  They then extrapolated these masses to the one corresponding to
the physical pion mass to obtain a chiral monopole
mass value of $0.727\pm 0.016$ GeV.  The (monopole) form factor calculated
with that mass (including its uncertainty) is indicated by the shaded band in
Fig. \ref{fig:latdisp}, cut off at the highest \qsq\ point of the
lattice simulation.  This result begins to trend away from the $Q^2>1.5$
\gevsq\ experimental data.  It remains to be seen how these results would
be affected by our Sec.~\ref{sec:monopole} comments on the applicability of the
monopole parameterization in this \qsq\ range.

\subsection{Dispersion Relation with QCD Constraint}

Dispersion relations are based on constraints posed by causality and
analyticity, and relate the timelike and spacelike domains of the pion form
factor on the complex plane.  In
principle the technique is exact, but our incomplete knowledge of the
scattering amplitudes over the whole complex plane, and in particular the
incomplete understanding of the contribution of all of the poles in the
timelike region, creates uncertainties.  Authors address these uncertainties by
imposing additional constraints, such as the role of higher timelike resonances
like the $\rho'''$, or chiral perturbation constraints near the spacelike
threshold, or that \fpi\ must approach its expected asymptotic value at very
large \qsq\ \cite{donoghue, buck, geshkenbein, watanabe, melikhov}.  We compare
the \fpi\ data to the dispersion relation analysis of B.V. Geshkenbein
\cite{geshkenbein} in Fig. \ref{fig:latdisp}.  The displayed uncertainty band
is obtained by assuming different distributions of zeroes in the complex
$s$-plane.  This results in a band that grows with \qsq, with the
`no zeroes' curve lying nearly at the lower end of the band.  Our highest \qsq\
data lie above the `no zeroes' curve, but below the `improved maximum \fpi'
limit.

\subsection{QCD Sum Rules}

\begin{figure}[b]
\begin{center}
        \includegraphics[height=8.5cm,angle=90]{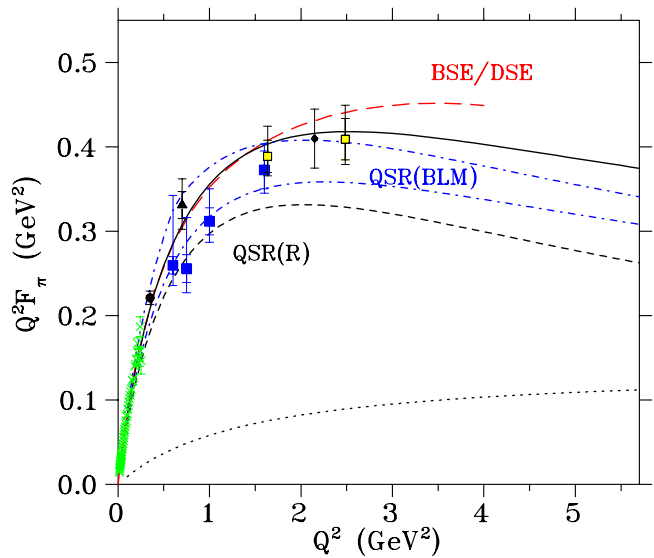}
\end{center}
\caption{(Color online)
The \fpi\ data of Fig. \ref{fig:q2fpi_mono} are compared with the QCD Sum Rules
calculations  of Refs. \protect{\cite{radyushkin91,brag07}} and the
Bethe-Salpeter equation
model utilizing dressed quark propagators via the Dyson-Schwinger equation
of Ref. \protect{\cite{maris00b}} [long dashed].  
For the calculation of Ref. \cite{radyushkin91}, three curves are shown:
[dotted] $F^{hard}_{\pi}$, [short-dashed] $F^{soft}_{\pi}$, and [solid] the sum
$F^{soft}_{\pi}+F^{hard}_{\pi}$.  For the calculation of Ref. \cite{brag07},
two dot-dashed curves are shown: 
[lower] $s_0=\frac{4\pi^2f_{\pi}^2}{1+\alpha_s(Q^2)/\pi}$, 
[upper] $s_0=0.65$ \gevsq.  
\label{fig:qsrdse}}
\end{figure}

The QCD sum rule approach is designed to interpolate between the perturbative
and non-perturbative sectors using dispersion relation methods in combination
with the operator-product expansion.  While the practical implementation of
this approach cannot claim to be rigorously derived from QCD, its intuitive
value is that it provides a bridge between the low- and high-energy properties
of QCD \cite{thomas}.  A number of authors have applied this technique with
good success to the pion form factor \cite{brag07,braun,nesterenko,
radyushkin91,forkel}.  In the calculation of Radyushkin \cite{radyushkin91},
QCD sum rules were used to give a local quark-hadron duality estimate of the
soft wave function 
\begin{equation}
F^{soft}_{\pi}(Q^2)=1-\frac{1+6 s_0/Q^2}{(1+4s_0/Q^2)^{3/2}},
\label{eqn:qsr_soft}
\end{equation}
where the duality interval, $s_0$, which within the QCD sum rule approach is
determined by the magnitude of the quark and gluon condensates, was taken as
$4\pi^2f_{\pi}2\approx 0.7$ \gevsq.  This soft
calculation, shown in Fig. \ref{fig:qsrdse}, under-estimates the
data by about 25\%.  For the hard contribution, a simple model based on the
interpolation between the behavior near $Q^2=0$ (related by the Ward identity
to the
$O(\alpha_s)$ term of the 2-point correlator) and the asymptotic behavior was
used
\begin{equation}
F^{hard}_{\pi}(Q^2)=\frac{\alpha_s}{\pi}\frac{1}{(1+Q^2/2s_0)}.
\end{equation}
The sum, $F^{soft}_{\pi}+F^{hard}_{\pi}$, is in excellent agreement with the
data.

More recently, Braguta, Lucha and Melikhov \cite{brag07} have replaced the
simple ansatz leading to Eqn. \ref{eqn:qsr_soft} with an expression including
explicit corrections up to $O(\alpha_s)$.  Since the higher-order
corrections needed to apply these results with authority to the intermediate
\qsq\ region are beyond the capacity of their two-loop calculation, there is a 
model dependence in their numerical result,
which is reflected in the two different curves for $s_0=0.65$ \gevsq\ and
$s_0=\frac{4\pi^2f_{\pi}^2}{1+\alpha_s(Q^2)/\pi}$ shown in Fig. \ref{fig:qsrdse}.

\subsection{Bethe-Salpeter Equation}

The Bethe-Salpeter equation (BSE) is the conventional formalism for the
treatment of relativistic bound states.  In this formalism, a meson is
described by a covariant wavefunction, which depends on the four momenta of its
constituent quarks.  Although formally correct, complications arise as the
interplay between different configurations, such as $q$-$\bar{q}$ and
$q$-$\bar{q}$-$g$, are implicitly buried in the potential and scattering
amplitudes used in analyzing hadronic processes, and as a result, these
potentials and scattering amplitudes are nearly intractable.  The light-front
Bethe-Salpeter model is a means to handle this problem by breaking the BSE into
separate hard and soft components.  A variety of models incorporating a
confining potential which dominates at low \qsq, and a QCD-based interaction
which dominates at high \qsq, are given in
Refs. \cite{munz,melo02,melo06,szczurek,kisslinger01}.

Another approach is to use the Dyson-Schwinger equation (DSE) to obtain
dressed quark propagators which may be used in the solution of the BSE.  The
Dyson-Schwinger approach to nonperturbative QCD has many advantages.  It is
consistent with quark and gluon confinement, it automatically generates
dynamical chiral symmetry breaking, and the solution is Poincare invariant.  In
the work of Maris, Tandy, and Roberts, the meson Bethe-Salpeter amplitudes and
quark-photon vertex are obtained as solutions of the homogeneous and
inhomogeneous BSE, and the dressed quark propagators are obtained from the
quark DSE.  The model parameters are fixed by requiring $f_{\pi}$ and $m_{\pi}$
to be in good agreement with the data \cite{maris98} and then $r_{\pi}$ and 
\fpi\ are predicted with no further adjustment of parameters
\cite{maris00a,maris00b}.  Their calculation is shown in Fig. \ref{fig:qsrdse}.
It is in excellent agreement with our data up to \qsq=1.60 \gevsq.  To extend the
validity of the model to higher \qsq, a more complete description that takes
meson loop corrections into account self-consistently is required
\cite{maris00b}.

\subsection{Local Quark-Hadron Duality}

Quark-hadron duality relations link the hadronic structure information
contained in exclusive form factors and inclusive structure functions by making
strong assumptions of locality \cite{melentkep}.  
While local quark-hadron duality is an
expected consequence of QCD at asymptotically large momenta, it is not at all
clear how well it could work at finite \qsq\ \cite{isgur01}.  
And if it does, it may be due to accidental cancellations of higher twist effects.
Nevertheless, it is worthwhile to compare predictions based on
quark-hadron duality with the measured data, especially since duality is
expected to work better at higher \qsq, in contrast to many other approaches.

The approximate relationship between the pion elastic form factor and the pion
structure function $F_2^{\pi}=\nu W_2^{\pi}$ was found by Moffat and Snell
\cite{moffat},
\begin{equation}
[\fpi (\qsq)]^2\approx \int_1^{\omega_{max}} F_2^{\pi}(\omega) d\omega,
\label{eqn:moffat}
\end{equation}
where $\omega=1/x$, and
the upper limit of integration is chosen to select the elastic
contribution to the inclusive structure function.  
In applying this formula use is made of the Drell-Yan-West \cite{drell-yan,
west} relation, which is based on a field-theoretic parton model that predates QCD. 
It predicts that if the asymptotic
behavior of a form factor is $(1/\qsq)^n$, the corresponding structure function
should behave as $(1-x)^{2n-1}$ as $x\rightarrow 1$.  This leads to the
prediction $F_2^{\pi}(x\rightarrow 1)\sim (1-x)$.

\begin{figure}
\begin{center}
        \includegraphics[height=8.5cm,angle=90]{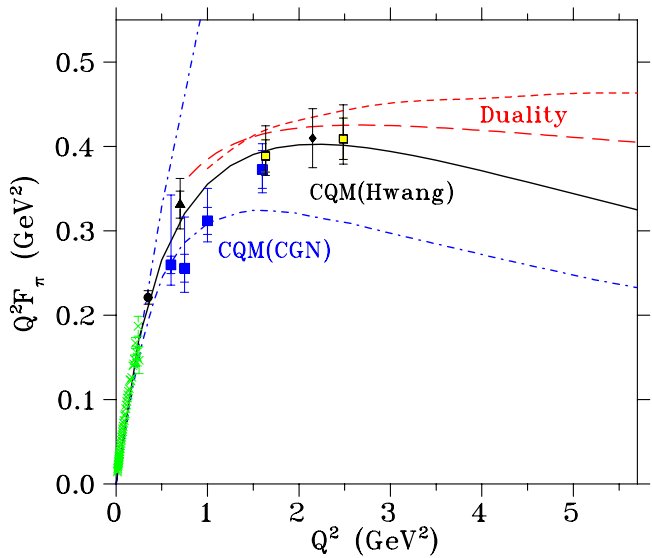}
\end{center}
\caption{(Color online)
The \fpi\ data of Fig. \ref{fig:q2fpi_mono} are compared with the local
quark-hadron duality analysis of W. Melnitchouk \protect{\cite{mel03,mel06}},
and the constituent quark model calculations of Refs. \protect{\cite{car94,hwang01}}.
For the duality calculation, two curves are shown: [short-dashed] leading-order
analysis of Ref. \protect{\cite{mel03}}, [long-dashed] next-to-leading order analysis
of Ref. \cite{mel06}.
For the quark model calculations by Cardarelli {\em et al.} \protect{\cite{car94}},
two curves are shown: [upper dot-dashed] point-like quarks, [lower dot-dashed] 
quarks with a monopole form factor.
\label{fig:ducqm}}
\end{figure}

The existence of Drell-Yan $F_2^{\pi}$ data allows a quantitative test of
Eqn. \ref{eqn:moffat} using only phenomenological input.  Calculations
\cite{mel03, mel06} based on the leading-order analysis of $F_2^{\pi}$ data by
Ref.  \cite{fnal}, and the next-to-leading order analysis of
Ref. \cite{wije05}, are shown in Fig. \ref{fig:ducqm}.  In both cases, the
magnitude of the \fpi\ prediction is dependent on the value chosen for
the inelastic cutoff $\omega_{max}$ (and corresponding $W_{max}$) 
in Eqn. \ref{eqn:moffat}.  Local duality is
expected not to work at lower \qsq.  This is reflected in
the poor description of the $Q^2<1$ \gevsq\ form factor data.  However,
above $Q^2>2$ \gevsq, the next-to-leading order analysis is consistent
with our data.

\subsection{Constituent-Quark Model}

There are many \fpi\ calculations using a variety of constituent-quark models
\cite{anisovich, cardarelli95, allen98, choi99, car94, hwang01, krutov01, amghar,
krutov02, sengbusch, coester05}.  The differences in approach
typically involve differences in the treatment of the quark wave functions, or
the inclusion of relativistic effects.  Fig. \ref{fig:ducqm} shows the result
of calculations by Cardarelli {\em et al.} \cite{car94} and by Hwang
\cite{hwang01}.  Both are relativistic quark models on the light front.
Ref. \cite{car94} uses the effective $q\bar{q}$ Hamlitonian of \cite{god85}, 
which contains a one-gluon-exchange term and a linear confining term, and which
describes a large set of meson spectroscopic data.  Use of
this interaction results in large high-momentum components, and \fpi\ is
strongly overpredicted (upper dot-dashed curve in Fig.~\ref{fig:ducqm}).  
This can be
cured in a way that is consistent with the notion of a constituent quark, by
assuming a form factor for the latter.  Taking a monopole form for the latter
and adjusting the mass parameter so that the measured pion charge radius is
reproduced, results in the lower dot-dashed curve shown.

The model of Ref. \cite{hwang01} allows a consistent and fully
relativistic treatment of quark spins and center-of-mass motion to be carried
out.  A power-law wave function is used, whose parameters are determined from
experimental data on the charged pion decay constant, the neutral pion
two-photon decay width, and the charged pion electromagnetic radius.  The
charge and transition form factors of the charged pion and the branching ratios
of all observed decay modes of the neutral pion are then predicted.  The
calculation is in very good agreement with our \fpi\ data.

Li and Riska \cite{li07} asked if the empirical \fpi\ data exclude the presence
of a significant sea-quark configuration in the charged pion.  They performed a
constituent quark model calculation which was extended to include explicit
sea-quark components in the pion wave function.  They found that these
sea-quark contributions grew with increasing \qsq, because they allowed the
momentum transfer to be shared by a greater number of constituents, and so were
less-suppressed at high \qsq\ than configurations which involved only a
$\bar{q}q$ pair.  Although their analysis was model-dependent, they found that
our data allowed an approximate $20\pm 20\%$ sea-quark component, with the data
point at \qsq=2.45 \gevsq\ providing the greatest constraint.

\subsection{Holographic QCD}

A recent theoretical development is the AdS/CFT correspondence \cite{mal98}
between weakly-coupled string states defined on a 5-dimensional anti-de Sitter
space-time ($\rm AdS_5$) and a strongly-coupled conformal field theory (CFT) in
physical space-time.  The goal of holographic QCD models is to find a
weakly-coupled theory for which the dual strongly-coupled theory is as close to
QCD as possible, and so allow analytic solutions of hadronic properties in the
non-perturbative regime to be performed.  In these models, confinement is
simulated by imposing boundary conditions on the extra fifth dimension $z$
\cite{pol02}.  In the ``hard-wall'' variant, confinement is modeled by a hard
cutoff at a finite value $z=z_0=1/\Lambda_{QCD}$.  This has the advantage of
simplicity but produces the unphysical Regge trajectory $M_n^2\sim n^2$.  The
``soft-wall'' variant replaces the hard-wall boundary with an oscillator-type
potential, and produces the more phenomenologically realistic Regge behavior
$M_n^2\sim n$.

\begin{figure}
\begin{center}
        \includegraphics[height=8.5cm,angle=90]{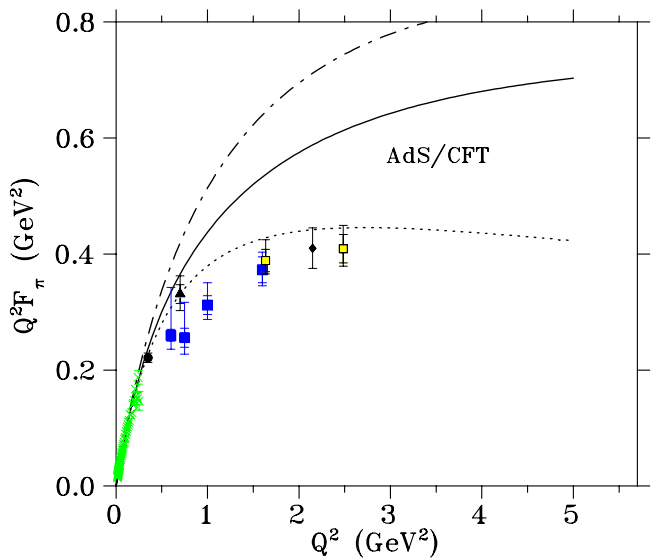}
\end{center}
\caption{(Color online)
The \fpi\ data of Fig. \ref{fig:q2fpi_mono} compared with the holographic QCD
model calculations by H.J. Kwee and R.F. Lebed \protect{\cite{kwe07}}.  The
curves are: [solid] ``hard-wall'' and [dot-dash] ``soft-wall'', both with
parameters fit to $m_{\pi}$, $m_{\rho}$ and $f_{\pi}$, and [dash] ``soft-wall''
with $\sigma=262$ MeV to improve the fit to \fpi\ at higher \qsq\ but
destroying the agreement with the other observables.
\label{fig:adscft}}
\end{figure}

Several authors have applied holographic models to the pion form factor
\cite{brod07,kwe07,kwe07b,grig07}.  Complications arise when one introduces
spontaneous and explicit chiral symmetry breaking effects
into the soft-wall holographic QCD model.  Refs. \cite{kwe07,grig07} take
different approaches to this problem.  Grigoryan and Radyushkin \cite{grig07}
consider only the hard-wall variant, and then estimate a soft-wall correction
from their previous vector meson study \cite{grig07a}.  They conclude that a
full analytic calculation would likely follow the \fpi\ data only in the
$Q^2<1$ \gevsq\ region, while overshooting it above $Q^2\sim 2$ \gevsq.  
The calculations by Kwee and Lebed
\cite{kwe07,kwe07b} are numerical.  Both the hard-wall and the soft-wall
calculations predict charge radii that are too small, especially for the
soft-wall case (see Fig. \ref{fig:adscft}).  By allowing the parameters of the
soft-wall model (originally fixed by $m_{\rho}$, $m_{\pi}$, and $f_{\pi}$) to
vary, they find it is possible to describe \fpi\ at either high \qsq\ or low
\qsq, but not both.  Issues in ongoing discussions \cite{kwe07b,grig07}
on the AdS/CFT approach
include the applicability of this model to the larger \qsq\ region where
partonic degrees of freedom become appreciable, and the treatment of chiral
symmetry breaking.

\section{Summary and Outlook}

Values for the charged pion form factor, $F_{\pi}(Q^2)$, have been extracted
for \qsq=0.60-2.45 \gevsq\ from the longitudinal cross sections
$\sigma_{\mathrm{L}}(t)$ for the $^1$H$(e,e^{\prime}\pi^+)n$ reaction recently
measured at JLab.  \fpi\ values were also extracted from older experimental data
acquired at DESY.  The Cornell data are not included in this analysis because
these \sigl\ were not obtained in a true L/T-separation, but instead by
subtracting a certain assumption for \sigt.
In addition, the higher \qsq\ data have excessively large values of $-t_{min}$.

The form factor extraction requires the use of a model
incorporating both the $\pi^+$ production mechanism as well as the effect of
the nucleon.  Several approaches to extract \fpi\ from the data, including the
Chew-Low extrapolation method, various types of Born Term Models, and newer
models utilizing Regge trajectories and effective Lagrangians, were reviewed.
By using specially generated test data, it was found that extrapolating to the
pole at $t=m_\pi^2$, as is done in the Chew-Low method, cannot be used in
practice, because there is no way to determine the order of the polynomial to
use for the extrapolation, and because even small uncertainties in the measured
cross sections lead to a large uncertainty in \fpi.

From the models available for determining \fpi\ from the measured values of
\sigl, the VGL Regge model~\cite{van97} was chosen, since it contains no
ad hoc parameters, and its validity has been well established over a wide
kinematic range in $t$ and $W$ for both electroproduction and photoproduction data.
The VGL
model gives a rather good description of both the $t$ and the $W$ dependence of
the JLab data at values of \qsq$>1$ \gevsq, but especially at $Q^2=0.60$
\gevsq\ the fall-off of the data with $-t$ is steeper than that of the model.
In the cases where the VGL model described well the $t$-dependence of the
\sigl\ data, the value of \fpi\ was determined by fitting the model to the
data.  Otherwise, the value of \fpi\ was determined by extrapolating the fit of
the model to $t=t_{min}$.  An additional `model uncertainty' has been estimated
by using two different assumptions for an interfering background that could be
responsible for this discrepancy between the data and VGL model.  The fact that
the discrepancy, and hence the model uncertainty, is very small at higher
values of \qsq\ and $W$ suggests that effects from nucleon resonances play a
role in the data at lower \qsq\ and $W$.

It is stressed that the cross sections are the actual observables measured by
the experiment, and that the extracted values of \fpi\ are inherently dependent
on the model used to extract them.  The development of additional models for
the $^1$H$(e,e^{\prime}\pi^+)n$ reaction would allow further exploration of the model
dependence of the extraction of \fpi\ from the same cross section data.  On the
experimental front, proposed measurements \cite{fpi12} after the completion of the JLab
upgrade are expected to better establish the validity of any used
model by investigating, for example, the $W$-dependence of the results.

The results for \fpi, extracted from our data and from the DESY data with the
use of the VGL model, are presented together with their experimental and model
uncertainties.  Above $Q^2\approx 1.5$ \gevsq, these data are systematically
below the monopole parameterization based on the empirical pion charge radius.
The data are also compared to a selection of 
calculations, including those based on pQCD,
Lattice QCD, Dispersion Relations, QCD Sum Rules, Bethe-Salpeter Equation,
Local Quark-Hadron Duality, Constituent-Quark Model, and Holographic QCD.
There has been tremendous progress in the theory of hadronic structure physics
in the past decade, as evident by the many new approaches under development.
However at present, the intermediate \qsq\ regime remains a significant
challenge.  Several different approaches concur that up to at least \qsq=2.5
\gevsq, the \fpi\ data are far above the estimated `hard' (perturbative)
contribution, and that `soft' (non-perturbative) contributions likely dominate
in this region.  Data expected to be taken \cite{fpi12} after the completion of
the JLab upgrade, up to at least $Q^2=6.0$ \gevsq, are expected to indicate 
whether the higher-twist mechanisms
dominate \fpi\ until very large momentum transfer, or not.

\section{Acknowledgments}
The authors thank Drs. Guidal, Laget, and Vanderhaeghen for
stimulating discussions and for modifying their computer program for our needs.
We also thank Dr. Obukhovsky for supplying the result of their
model calculations and for many informative discussions.  This work is
supported by DOE and NSF (USA), NSERC (Canada), FOM (Netherlands), NATO, and
KOSEF (South Korea).

\end{document}